\documentclass[12pt]{iopart}
\usepackage[]{graphicx}
\begin{document}

\title[Manuscript preparation guidelines]
{Coupled ferro-antiferromagnetic Heisenberg bilayers investigated by
many-body Green's function theory}

%\title[Author guidelines for IOPP journals]{Preparing an article for
%publication in an Institute of Physics Publishing journal using \LaTeXe}
\author[P. Fr\"obrich]{P Fr\"obrich
\footnote{also at Institut f\"ur Theoretische Physik,
Freie Universit\"at Berlin, Arnimallee 14, D-14195 Berlin, Germany}
\footnote{Corresponding
author: e-mail: {\sf froebrich@hmi.de}, Phone:+49\,30\,8062\,2684}
and P J Kuntz}
\address{
Hahn-Meitner-Institut Berlin, Glienicker Stra{\ss}e 100,
D-14109 Berlin, Germany}
\author[P J Jensen]{P J Jensen}
\address{Institut f\"ur Theoretische Physik,
Freie Universit\"at Berlin, Arnimallee 14, D-14195 Berlin, Germany}
%\author{Neil Scriven\dag\ and Romneya Robertson\ddag
%\footnote[3]{To
%whom correspondence should be addressed (romneya.robertson@iop.org)}
%}
%
%\address{\dag\ Production Editor, Institute of Physics
%Publishing, Dirac
%House, Temple Back, Bristol BS1 6BE, UK}
%
%\address{\ddag\ Electronic Services Specialist, Institute of Physics
%Publishing, Dirac House, Temple Back, Bristol BS1 6BE, UK}
%%
%% Information for the second author
%\author[P.J. Kuntz]{P.J. Kuntz\inst{1}}
%%\footnote{Second author footnote.} may be inserted after the name.
%%
%% Information for the third author
%\author[M. Stutzmann]{Martin Stutzmann\inst{1,2}}
%%\footnote{Third author footnote.} may be inserted after the name.
\begin{abstract}
A theory of coupled ferro-and antiferromagnetic Heisenberg layers is developed
within the framework of a many-body Green's function theory (GFT) that
allows non-collinear magnetic arrangements by introducing
sublattice structures. As an example,
the coupled ferro- antiferromagnetic (FM-AFM) bilayer is investigated. We
compare the
results with those of  bilayers with purely ferromagnetic or antiferromagnetic
couplings. In each case
we also show the corresponding results of mean field theory (MFT), in which
magnon excitations are completely neglected. There are significant differences
between GFT and MFT.
A remarkable finding is that for the coupled FM-AFM bilayer the critical
temperature decreases with increasing interlayer coupling strength for a simple
cubic lattice, whereas the opposite is true for an fcc lattice as well as for
MFT for both lattice types.

 \end{abstract}

%% maketitle must follow the abstract.
%\maketitle                   % Produces the title.

\pacs{75.10.Jm, 75.70.Ak}


\maketitle

%%    Give a maximum of six PACS code in numerical order.
%[\it Please insert a maximum of 6
%relevant codes, see \sf www.aip.org/pacs.]}

%% \pretitle{Editor's Choice}

%% We have a short and a long form for the title. The short form
%% (optional argument) goes into the running head.

%% Please do not enter footnotes or \inst{}-notes into the optional
%% argument of the author command. The optional argument will go into
%% the header.  If there is only one address the marker \inst{x} may be
%% omitted.

%% Information for the first author.
%%    \dedicatory{This is a dedicatory.}
%% If there is not enough space inside the running head
%% for all authors including the title you may provide
%% the leftmark in one of the following three forms:

%% \renewcommand{\leftmark}
%% {First Author: A Short Title}

%% \renewcommand{\leftmark}
%% {First Author and Second Author: A Short Title}
\def\la{\langle}

\def\ra{\rangle}

\def\K{\mathord{\cal K}}

\def\ltsim{\mathop{\,<\kern-1.05em\lower1.ex\hbox{$\sim$}\,}}

\def\gtsim{\mathop{\,>\kern-1.05em\lower1.ex\hbox{$\sim$}\,}}

\renewcommand{\leftmark}
{P Fr\"obrich, P J Kuntz, P J Jensen: Coupled ferro-antiferromagnetic
bilayers...  }
%

%xxxxxxxxxxxxxxxxxxxxxxxxxxxxxxxxxxxxxxxxxxxxxxxxxxxxxxxx
\section{Introduction}

In the interface region of coupled ferro- and antiferromagnetic systems
there is a magnetic reordering known as the magnetic proximity effect (MPE).
MPE has for a long time attracted the interest of researchers
\cite{MPE} since the
constituents show a novel magnetic arrangement different from that of
the bulk. The interest in such
interfaces has revived lately with respect to the
exchange bias effect \cite{rev-EB}, which occurs when a thin
ferromagnetic (FM) film is deposited on an antiferromagnetic (AFM)
material. If the latter is an `in-plane AFM',  and the number of bonds between parallel and
antiparallel spin pairs
across the interface is the same, the interface is `compensated'.
In this case the AFM
often assumes an almost orthogonal magnetization with respect to
the FM magnetic direction, while the spins of the AFM interface layer
reside in a `spin-flop-phase', in analogy to an
AFM system in an external magnetic field \cite{neel}.

In previous studies investigating FM-AFM interfaces
the magnetization of each FM layer is usually considered to be
collinearly ordered and to rotate as a whole \cite{koon,SRT}.
Results concerning the spin reorientation transition (SRT)
have been obtained for various magnetic systems but,
to the best of our knowledge, only in presence of magnetic
anisotropies \cite{SRT}.
It is important to stress that, although we consider in the present work
anisotropic interactions too, a noncollinear magnetization
in coupled FM-AFM bilayers is caused mainly by isotropic exchange
interactions. Whereas previous work describes a net magnetic
reorientation of the total system, we consider the reorientation to
take place in  magnetic sublattices, leaving
the net magnetic orientation of each layer almost unchanged.
In the case of a compensated FM-AFM interface,
the MPE extends to the FM layers close to the interface.
Then, the magnetic structure of each FM layer is represented, in
perfect analogy to the AFM layers, by two juxtaposed sublattices
with different but uniform magnetization directions. Allowing a
nonuniform intralayer magnetic structure in the FM subsystem leads to
new features, which in turn are strongly dependent on the underlying
lattice symmetry.

Often for simplicity, a mean field theory (MFT) is
used to describe the magnetic reorientation.
In a recent MFT \cite{JKD} study of a coupled FM-AFM system for both sc(001)
and fcc(001) lattices, a variety of different magnetic
configurations emerge, depending on the parameter values.  Usually the
subsystem with the larger
ordering temperature induces a magnetic order in the other one (MPE).
For coupled sc(001) systems, both FM and AFM films are perturbed from
their collinear magnetic order and exhibit similar behavior.
This symmetry is absent for fcc(001) films, which, under certain
circumstances, may exhibit two different critical temperatures.
An advantage of MFT is that many results can be derived  analytically for
simple bilayer systems.

We stress that for thin magnetic films,  collective
magnetic excitations (spin waves) are particularly important. These are
neglected completely by MFT.
Such excitations are taken into account, for example, by many-body
Green's function theory. There is a large amount of work applying this
theory to
thin ferromagnetic films, in particular in connection with the reorientation of
the magnetization as a function of temperature and film thickness. We
mention only a few papers that cite further literature \cite{FJKE00,
FK03, WDFJK04, SKN05}. Antiferromagnetic films have also been treated, e.g. in
\cite{D91,WHQW04}. Not as much work has been done in which Green's
function theory treats the coupling of ferromagnetic layers
to antiferromagnetic layers: in reference \cite{MUH93}, a
bilayer is investigated and
reference \cite{MU93} treats an extension to multilayers. In both cases, only a
collinear magnetization is considered. In reference \cite{WD04}, a
ferromagnetic film is coupled to an antiferromagnetic layer; however, the
orientation
of the magnetization of the antiferromagnet is frozen. Other work considers an
antiferromagnetic coupling between ferromagnetic layers \cite{MU94,L04,NNDN04}.

The new feature of the present work is to allow  an in-plane
reorientation of the magnetizations of the ferro- and antiferromagnetic layers
due to the interlayer coupling as in the MFT approach of \cite{JKD} but using
Green's function theory.
We restrict ourselves to Heisenberg systems with spin $S=1/2$ with an exchange
anisotropy. We do not consider this an essential restriction, because we
have shown in references \cite{FK04, FK104} for ferromagnetic layers
that through an appropriate choice of anisotropy parameters the  exchange- and
single-ion anisotropies yield very similar results, and that an appropriate
scaling leads to universal magnetization curves for different spin quantum
numbers. In the present paper, we examine in detail the magnetic
arrangement of the simplest system: a perfectly ordered bilayer
consisting of a FM monolayer that is coupled to a  AFM monolayer.
Thus, we do not address the exchange bias effect directly,
since this effect is  most likely
related to a certain amount of interface disorder \cite{EBdis}.

The paper is organized as follows. In Section 2 we develop the Green's function
theory and discuss the method of solution for the resulting equations.
In Section 3 we present numerical results for
bilayers with purely ferromagnetic or antiferromagnetic coupling as well as for
the coupled FM-AFM
bilayer. In particular the effect of the interface coupling $J_\mathrm{int}$
on the characteristics and magnitude of the MPE at zero and
finite temperatures is investigated. The resulting magnetic
arrangements for various kinds of bilayer systems and for their
corresponding ordering temperatures are determined.
In Section 4 we summarize the essential results and end with some remarks
concerning further development.

\section{The Green's function formalism for coupling ferro- and
antiferromagnetic layers}

The starting point is a XXZ-Heisenberg Hamiltonian consisting of an
isotropic Heisenberg exchange
interaction with strength $J_{ij}$ between nearest neighbour lattice sites,
exchange (non-localized)
anisotropies in the $x$- or $z$-directions having strengths $D^x_{ij}$ and
$D^z_{ij}$, respectively, and an external magnetic
field ${\bf B}=(B^x,0,B^z)$ confined to the film plane, which is the
$xz$-plane:
\newpage
\begin{eqnarray}
{\cal H}=&-&\frac{1}{2}\sum_{<ij>}J_{ij}(S_i^-S_j^++S_i^zS_j^z)
-\frac{1}{2}\sum_{<ij>}(D^x_{ij}S_i^xS_j^x+D^z_{ij}S_i^zS_j^z)
\nonumber\\
&-&\sum_k\Big(B^xS_i^x+B^zS_k^z\Big).
\label{1}
\end{eqnarray}
Here the notation $S_i^{\pm}=S_i^x\pm iS_i^y$  is
introduced, and $<ij>$ indicates
summation over nearest neighbours only, where $i$ and $j$ are lattice site
indices. Because there is no field $B^y$ perpendicular to the film plane,
only a reorientation of the magnetization in the $xz$-plane is allowed.
For the FM-AFM bilayer we use $D_{ij}^z$ in the ferromagnetic layer and
$D_{ij}^x$ in the antiferromagnetic layer.

In this paper we restrict ourselves to systems with spin quantum number
$S=1/2$ and to a simple cubic (sc) lattice.
For this case we need the commutator Green's functions
\begin{equation}
G_{ij}^{\alpha -}(\omega)=\la\la
S_i^\alpha;S_j^-\ra\ra_\omega,
\label{2}
\end{equation}
where $\alpha=(+,-,z)$ takes care of all directions in space.
A generalization to spin quantum numbers $S>1/2$ is straight-forward by
introducing $G_{ij}^{\alpha,mn}=\la\la S^\alpha_i;(S_j^z)^m(S_j^-)^n\ra\ra$
with $m+n\leq 2S+1\ (m\geq 0;\ n\geq 1;\  m,n\  {\rm integer})$ \cite{FJKE00}.

The equations of motion for the Green's functions in the energy
representation are
\begin{equation}
\omega G_{ij}^{\alpha -}(\omega)=A_{ij}^{\alpha -}+\la\la
[S_i^\alpha,{\cal H}];S_j^-\ra\ra_{\omega}
\label{3}
\end{equation}
with the inhomogeneities
\begin{equation}
A_{ij}^{\alpha -}=\la[S_i^\alpha,S_j^-]\ra=
\left(\begin{array}{c}
2\la S^z_i\ra \delta_{ij}\\
0\\
-\la S^x_i\ra \delta_{ij}
\end{array}
\right),
\label{4}
\end{equation}
where $\la ...\ra={\rm Tr}(...e^{-\beta{\cal H}})/{\rm Tr}(e^{-\beta{\cal H}})$
denotes the thermodynamic expectation value.

In order to obtain a closed system of equations, the higher-order Green's
functions on the right hand sides are decoupled by a
generalized Tyablikov- (RPA) decoupling \cite{FJKE00}
\begin{equation}
\la\la S_i^\alpha S_k^\beta;S_j^-\ra\ra_\eta \simeq\la
S_i^\alpha\ra
G_{kj}^{\beta-}+\la S_k^\beta\ra G_{ij}^{\alpha-} .
\label{5}
\end{equation}
After introducing two sublattices per layer, which is necessary when dealing
with antiferromagnets, the resulting equations are Fourier transformed to
momentum space by
\begin{eqnarray}
G^{\alpha-}_{mn}({\bf k})&=&\frac{2}{N}\sum_{i_mj_n}G^{\alpha-}_{i_mj_n}
e^{-i{\bf k}({\bf R}_{i_m}-{\bf R}_{j_n})}, \nonumber\\
G^{\alpha-}_{i_mj_n}&=&\frac{2}{N}\sum_{\bf k}G_{mn}^{\alpha-}({\bf k})
e^{i{\bf k}({\bf R}_{i_m}-{\bf R}_{j_n})},
\label{6}
\end{eqnarray}
where $i_m,\ j_n$ are lattice site indices on the sublattices $m,n$, and $N$
is the number of lattice sites in the whole system. One obtains
\newpage
\begin{eqnarray}
\omega G_{mn}^{\pm-}&=&\left(\begin{array}{c}2\la S^z_m\ra\delta_{mn}\\ 0
\end{array}\right)\nonumber\\
& &\pm\Big(B^z+\sum_p\la S^z_p\ra(J_{mp}({\bf
0})+D_{mp}^{z}({\bf 0}))\Big)G_{mn}^{\pm-}
\nonumber\\ & &\mp \la S_m^z\ra\sum_p(J_{mp}({\bf k})+\frac{1}{2}D_{mp}^x({\bf
k}))G_{pn}^{\pm-} \nonumber\\
& &\mp\frac{1}{2}\la S^z_m\ra\sum_p D^x_{mp}({\bf k})G^{\mp-}_{pn}\nonumber\\
& &\mp\Big(B^x+\sum_p\la
S^x_p\ra(J_{mp}({\bf 0})+D_{mp}^{x}({\bf 0}))\Big)G_{mn}^{z-}\nonumber\\
& &\pm\la S^x_m\ra\sum_p(J_{mp}({\bf k})+D_{mp}^z({\bf
k}))G_{pn}^{z-},\nonumber\\
\omega G_{mn}^{z-}&=&-\la S^x_m\ra\delta_{mn}
\nonumber\\
& &-\frac{1}{2}\Big(B^x+\sum_p\la
S^x_p\ra(J_{mp}({\bf 0})+D^{x}_{mp}({\bf 0}))\Big)G_{mn}^{+-}\nonumber\\
& &+\frac{1}{2}\la S^x_m\ra\sum_p J_{mp}({\bf k})G^{+-}_{pn}\nonumber\\
& &+\frac{1}{2}\Big(B^x+\sum_p\la
S^x_p\ra(J_{mp}({\bf 0})+D_{mp}^{z}({\bf 0}))\Big)G_{mn}^{--}\nonumber\\
& &-\frac{1}{2}\la S^x_m\ra\sum_pJ_{mp}({\bf k})G_{pn}^{--}.
\label{7}
\end{eqnarray}
For a square lattice with lattice constant $a_0=1$, one has
four nearest-neighbour {\em intralayer} couplings with sublattice
indices $n,m$ from the same layer
\begin{equation}
\begin{array}{ll}
J_{mn}(\mathbf{0})=q_0\,J_{mn} \;, &
J_{mn}(\mathbf{k})=\gamma_0(\mathbf{k})\,J_{mn} \;, \\[0.2cm]
D_{mn}^{x,z}(\mathbf{0})=q_0\,D_{mn}^{x,z} \;, &
D_{mn}^{x,z}(\mathbf{k})=\gamma_0(\mathbf{k})\,D_{mn}^{x,z} \;,
\end{array}
\label{8}
\end{equation}
with the {\em intralayer} coordination number $q_0=4$ and
the momentum-dependent Fourier factor
\begin{equation} \label{9}
\gamma_0({\bf k})= 2(\cos k_x+\cos k_z) \;.
\end{equation}
Correspondingly, for the nearest neighbour {\em interlayer} couplings, with
$m,n$  now being sublattice indices from different layers, one obtains
\begin{equation} \label{10}
\begin{array}{ll}
J_{mn}(\mathbf{0})=q_\mathrm{int}\,J_\mathrm{int} \;, &
J_{mn}(\mathbf{k})=\gamma_\mathrm{int}(\mathbf{k})
\,J_\mathrm{int} \;, \\[0.2cm]
D_{m,n}^{x,z}(\mathbf{0})=q_\mathrm{int}\,D_\mathrm{int}^{x,z} \;, &
D_{mn}^{x,z}(\mathbf{k})=\gamma_\mathrm{int}(\mathbf{k})
\,D_\mathrm{int}^{x,z} \;.
\end{array}
\end{equation}
For sc stacking, the {\em interlayer} coordination number and the
corresponding Fourier factor are given by
\begin{equation} \label{11}
q_\mathrm{int}=\gamma_\mathrm{int}(\mathbf{k})=1 \;,
\end{equation}
which is assumed in the following calculations if not stated otherwise.

{\noindent}For  fcc or bcc stacking one has for comparison
\begin{equation}\label{11a}
q_\mathrm{int}=4\ \ \  {\rm and}\ \ \
\gamma_\mathrm{int}(\mathbf{k})=4\,\cos(k_x/2)\,\cos(k_z/2).
\end{equation}
The mean field approximation  is obtained by neglecting the Fourier factors,
i.e. $\gamma_0(\mathbf{k})=\gamma_\mathrm{int}(\mathbf{k})=0$.
By choosing the appropriate signs of the exchange interaction and the
exchange anisotropy coupling constants, one can treat ferromagnetic,
antiferromagnetic, and mixed systems with coupled FM and AFM layers.

The general formalism is valid for any number of layers and sublattices.
If $Z$ is the total number of sublattices  of the
system, the dimension of the set of equations (\ref{7}) is $3Z^2$. Because we
restrict
ourselves in the present paper to the investigation of the bilayer problem,
there are four sublattices, and the system of equations
(\ref{7}) is of dimension $48$ with a corresponding Green's function vector.
Closer inspection reveals that the system of equations has the following
substructure
\begin{equation}
\left( \omega {\bf 1}-\left( \begin{array}{cccc}
{\bf\Gamma} & 0 & 0 & 0 \\
0 & {\bf\Gamma} & 0 & 0 \\
0 & 0 & {\bf\Gamma} & 0 \\
0 & 0 & 0 & {\bf\Gamma}
\end{array}\right)\right)\left( \begin{array}{c}
{\bf G}_{1} \\ {\bf G}_{2} \\ {\bf G}_3 \\ {\bf G}_{4}
\end{array} \right)=\left( \begin{array}{c}
{\bf A}_{1} \\ {\bf A}_{2} \\ {\bf A}_3 \\
{\bf A}_{4} \end{array}
\right) \ ;
\label{12}
\end{equation}
where the diagonal blocks ${\bf \Gamma}$  are identical $12\times 12$
matrices, whose explicit form can be read off from equations (\ref{7}). The
sublattice Green's functions ${\bf G}_n\ (n=1,2,3,4)$ are vectors of dimension
$12$ consisting of $4$ subvectors, each of dimension $3$:
\begin{equation}
{\bf G}_n=\left(\begin{array}{c}
{\bf G}_{1n}\\ {\bf G}_{2n}\\ {\bf G}_{3n} \\{\bf G}_{4n}
\end{array}
\right),\ \ {n=1,2,3,4}\ ,
\label{13}
\end{equation}
where the 3-component vectors are
\begin{equation}
{\bf G}_{mn}=\left(\begin{array}{c}
{\bf G}_{mn}^{+-}\\ {\bf G}_{mn}^{--}\\ {\bf G}_{mn}^{z-}
\end{array}
\right), \ {m=1,2,3,4}\ .
\label{14}
\end{equation}
The inhomogeneity vectors have the same structure:
\begin{equation}
{\bf A}_n=\left(\begin{array}{c}
{\bf A}_{1n}\delta_{1n}\\ {\bf A}_{2n}\delta_{2n}\\ {\bf A}_{3n}\delta_{3n}
\\{\bf A}_{4n}\delta_{4n} \end{array}
\right),\ \ {\bf A}_{nm}=\left(\begin{array}{c}
2\la S^z_m\ra\\ 0 \\ -\la S^x_m\ra
\end{array}
\right), \ \ \ \ \ {m,n=1,2,3,4}\ .
\label{15}
\end{equation}
The big equation (\ref{12}) of dimension 48 for the bilayer can therefore be
replaced by 4 smaller equations of dimension 12:
\begin{equation}
(\omega{\bf 1}-{\bf \Gamma}){\bf G}_n={\bf A}_n\ \ \ \ {\rm for}\  n=1,2,3,4\
. \label{17}
\end{equation}

By invoking the spectral theorem, one obtains an expression for the diagonal
correlations in configuration space ${\bf C}_n$ (independent of ${\bf k}$) by
integrating over the correlations in momentum space ${\bf C}_n({\bf k})$
corresponding to the Green's functions ${\bf G}_n$ \cite{FJKE00}:
\begin{equation}
{\bf C}_n=\int {\bf C}_n({\bf k})=\int d{\bf k}{\bf R}\varepsilon{\bf L}{\bf
A}_n,\ \ \ \ \ n=1,2,3,4, \label{18}
\end{equation}
where ${\bf R}$ and ${\bf L}$ are matrices constructed from the right and left
eigenvectors of the eigenvalues of the {\em non-symmetric} matrix
${\bf \Gamma}$, and
$\varepsilon$ is a diagonal matrix with elements
$\delta_{ij}/(e^{\beta\omega_i}-1)$
obtained from eigenvalues $\omega_i\ (i=1, ..., 12)$ in the case
when all of them are not zero.

Unfortunately, equation (\ref{18}) cannot be used directly because
the $12\times 12 \ {\bf \Gamma}$-matrix turns out to have 4 zero eigenvalues.
These
have to be treated properly when applying the spectral theorem in order to
calculate the correlations. In reference \cite{FJKE00} we have shown that in
this case the correlations ${\bf C}_n({\bf k})$ obey the
equations
\begin{equation}
(1-{\bf R}^0{\bf L}^0){\bf C}_n({\bf k})={\bf R}^1\varepsilon^1{\bf L}^1{\bf
A}_n,  \ \ \ \  n=1,2,3,4\ ,
\label{19}
\end{equation}
where
\begin{equation}
{\bf R}=\big({\bf R}^0, {\bf R}^1\big)\ ,
\ \ \ \ \ \  {\bf L}=\left(\begin{array}{c}{\bf L}^0\\ {\bf L}^1
\end{array}\right)\ .
\label{20}
\end{equation}
The subscripts $0$ and $1$ label the matrices constructed from the right
and
left eigenvectors belonging to the zero and non-zero eigenvalues respectively,
and $\varepsilon^1$ is a diagonal $8\times 8$ matrix with elements
$\delta_{ij}/(e^{\beta\omega_i}-1)$
obtained from the non-zero eigenvalues $\omega_i\ (i=1, ..., 8)$.

As written, equation (\ref{19}) is formally correct but it still cannot be used
directly for two reasons: (1) The matrix $(1-{\bf R}^0{\bf L}^0)$ is
a projection operator on the non-null space and thus has no inverse; i.e.
we cannot isolate ${\bf C}_n({\bf k})$ unless $(1-{\bf R}^0{\bf L}^0)$ is
independent of ${\bf k}$. In the example of reference \cite{FK103}, this is the
case and one take the projector outside the integral.
This is not the case in the present example. (2)
Even if $(1-{\bf R}^0{\bf L}^0)$ were independent of ${\bf k}$, we still do not
know how to treat the non-diagonal correlations ${\bf C}_{mn}$ of a multilayer
problem.

We have considered these problems in recent publications \cite{FK103,FK204},
where we show that a solution is attained via the
the singular value decomposition (see e.g.
\cite{PFTV89}) of the ${\bf \Gamma}$-matrix
\begin{equation}
{\bf \Gamma}={\bf U}{\bf W}\tilde{\bf V}={\bf u}{\bf w}\tilde{\bf v},
\label{21}
\end{equation}
where ${\bf U}$ and $\tilde{\bf V}$ are orthogonal matrices, and ${\bf W}$ is
a diagonal matrix with the singular values on the diagonal, and ${\bf u}$ and
$\tilde{\bf v}$ are matrices obtained from ${\bf U}$ and $\tilde{\bf V}$ by
omitting the colums and rows corresponding to singular values zero.
Multiplying equation (\ref{19}) from the left by $\tilde{\bf v}$ and inserting
${\bf v}\tilde{\bf v}+{\bf v}_0\tilde{\bf v}_0=1$ gives
\begin{equation}
\tilde{\bf v}(1-{\bf R}^0{\bf L}^0){\bf C}_n({\bf k})={\bf
R}^1\varepsilon^1{\bf L}^1({\bf v}\tilde{\bf v}+{\bf v}_0\tilde{\bf v}_0)
{\bf A}_n\ ,
\label{22}
\end{equation}
and use of $\tilde{\bf v}{\bf R}^0=0, {\bf L}^1{\bf v}_0=0$, and
${\bf r}=\tilde{\bf v}{\bf R}^1$ and ${\bf l}={\bf L}^1{\bf v}$ leads to the
equations
\begin{equation}
\tilde{\bf v}{\bf C}_n({\bf k})={\bf r}\varepsilon^1{\bf l}
\tilde{\bf v}{\bf A}_n \ ,\ \ \ \ \ n=1,2,3,4\ .
\label{23}
\end{equation}
For more details of the formalism consult reference \cite{FK204}, where
we present  a systematic way of finding for each sublattice one
${\bf k}$-independent $\tilde{\bf v}$ vector having a layer
structure, i.e. $\tilde{\bf v}_n=(0,..,\tilde{v}_n,0,..,0)$. In this way
the non-diagonal correlations disappear from those rows in equation (\ref{23})
corresponding to $\tilde{\bf v}_n$, and the ${\bf k}$-integration can be
performed:
$\int d{\bf k}\tilde{\bf v}{\bf C}_n({\bf k})$=
$\tilde{\bf v}\int d{\bf k}{\bf C}_n({\bf k})=\tilde{\bf v}{\bf C}_n$. In the
present case
$\tilde{\bf v}_n$ is given by
\begin{eqnarray}\label{24}
\tilde{\bf v}_n
&=&\Big((\frac{1}{\sqrt{2}},-\frac{1}{\sqrt{2}},1)\delta_{1n},
(\frac{1}{\sqrt{2}},-\frac{1}{\sqrt{2}},1)\delta_{2n},\nonumber\\
& &(\frac{1}{\sqrt{2}},-\frac{1}{\sqrt{2}},1)\delta_{3n},
(\frac{1}{\sqrt{2}},-\frac{1}{\sqrt{2}},1)\delta_{4n}\Big)\ ,\ \ \
n=1,2,3,4.
\end{eqnarray}
Putting equation (\ref{24}) in equation (\ref{23}) yields 4 equations which
contain the
8 magnetization components implicitly. The necessary additional 4 equations are
obtained from the regularity conditions \cite{FJKE00}
\begin{equation}
\int d{\bf k}{\bf L}_0{\bf A}_n=\int d{\bf k}\tilde{\bf u}_0{\bf A}_n=0\ , \ \
\ \  n=1,2,3,4 , \label{25}
\end{equation}
which are obtained from the fact that the commutator Green's functions must be
regular at the origin, see e.g. \cite{Nol86}. The $\tilde{\bf u}_0$ are
constructed
from the null-eigenvectors of the singular value decomposition of the matrix
${\bf \Gamma}$. The resulting 8 integral
equations are solved self-consistently by the curve-following method described
in detail in the appendix of reference \cite{FKS02}.
Note that the $\tilde{\bf u}_0$ are determined numerically only
up to an orthogonal transformation. To ensure proper behaviour as a function of
${\bf k}$, $\tilde{\bf u}_0$ must be calibrated at each ${\bf k}$. A procedure
for effecting this is presented in an appendix of reference \cite{FK204}.
%\newpage

\section{Results}
In the following we present results for the bilayer ferromagnet, the bilayer
antiferromagnet, and the coupled ferro- and antiferromagnetic bilayer.
All calculations are for an in-plane orientation of the spins of both layers.
In each case we compare the results of Green's function theory (GFT) with those
of mean field theory (MFT) obtained by putting the
momentum-dependent terms equal to zero.

In order to see the effects of the interlayer coupling most clearly, we use
different exchange interaction strengths for both layers:

(a) FM-FM: $J_{1{\rm FM}}=100, J_{2{\rm FM}}=50$,

(b) AFM-AFM: $J_{1{\rm AFM}}=-100, J_{2{\rm AFM}}=-50$,

(c) FM-AFM: $J_{{\rm FM}}=100, J_{{\rm AFM}}=-50$.

Because of the Mermin-Wagner theorem \cite{MW66}, one needs anisotropies in the
Green's function description: We assume
$D^z=+1.0$ for FM layers and $ D^x=-1.0$ for AFM layers.  The
magnitude of the anisotropies is appropriate for 3d transition metal
systems.
For a compensated interface, the magnetizations of the FM and AFM layers
are almost orthogonal to each other even at $T=0$ because of the
interface exchange
interaction $J_{\rm int}$. We choose the FM magnetization to be oriented in the
$z$-direction and the AFM magnetization in the $x$-direction.
Our particular choice of the anisotropies supports this arrangement not
only at $T=0$ but also at finite temperatures. For other choices of
anisotropies the magnetic arrangement could be different.
The
interlayer coupling is assumed to be positive for the ferromagnetic bilayer
and negative for the
antiferromagnetic bilayer. For the coupled FM-AFM system, both signs
are used.
We consider three interlayer coupling constants with strength
$J_{\mathrm int}=30,\ 75,\ 160$, respectively, one smaller
than the weakest exchange interaction, one larger than the strongest exchange
interaction, and one in between.

\subsection{ The ferromagnetic and the antiferromagnetic bilayers}
Results for the FM and AFM bilayers are presented in this subsection.
We do so in order to have a basis for discussing the differences to the coupled
FM-AFM bilayer described later on.

In figure 1(a) we show the sublattice magnetizations of the ferromagnetic
bilayer
as a function of the temperature for three interlayer couplings calculated with
Green's function theory (GFT). The magnetization
profiles are different for the two layers (the magnetization is larger for the
layer with the larger exchange interaction)
but end in a common Curie temperature, which increases with
increasing strength of the interlayer coupling: $T_{\rm Curie}=50.66,\ 55.24,\
60.04$.

For the antiferromagnetic bilayer we use  the
same parameters as for the ferromagnetic bilayer except for a sign change. In
figure 1(b) we show the sublattice magnetizations of the antiferromagnetic
bilayer for two interlayer coupling strengths calculated with Green's function
theory. To
make the figures more transparent we leave out the result for
the intermediate
interlayer coupling strength. The corresponding magnetization curves lie in
between those of the other couplings.
One observes clearly the well known reduction of the
magnetizations
at low temperatures due to quantum fluctuations, which are missing
in MFT, see
figure 1(c). Since $|J_{\rm 1AFM}|>|J_{\rm 2AFM}|$ this reduction is larger for
the first
layer. With increasing temperature the magnetization curves of the
two layers
cross each other, a fact which was first observed by Diep \cite{D91}, and
finally end in a common N\'eel temperature. A larger
interlayer coupling leads
to a larger suppression  at low temperatures and to a larger N\'{e}el
temperature.
Whereas with the present choice of parameters the magnetization profiles of the
FM and AFM bilayers are rather
different at low temperatures, the critical temperatures turn out to be
identical: $T_{\rm Curie}=T_{\rm N\acute{e}el}$ (cf figures 1(a) and (1b)), a
fact that has already been discussed by Lines \cite{Lines64}.

For comparison, we show in figure 1(c)  the results of mean field theory
(MFT) with the same
parameters. The magnetization profiles as well as the critical temperatures are
identical for the ferromagnetic and antiferromagnetic bilayers.
As is well known, the Curie (N\'eel) temperatures
$(T_{\rm Curie (N\acute{e}el)}=102.10,\  107.25,\ 123.16\ $) are much larger
in MFT
due to the missing magnon excitations, with the present choice of parameters
by about a factor of 2.
Note that in MFT the Curie temperature
is not very sensitive to the anisotropies, as long as they are much smaller
than the exchange
interaction. In GFT, however, the
sensitivity is very much greater because of the
Mermin-Wagner theorem \cite{MW66} ($T_{\rm Curie (N\acute{e}el)}\rightarrow 0$
for $D^{z(x)}\rightarrow 0$). One observes also that the effect of the
interlayer
coupling on the
magnetization profiles is much stronger in MFT than in GFT.

\begin{figure}[htb]
\begin{center}
\protect
\includegraphics*[bb = 70 110 430 590,
angle=-90,clip=true,width=16cm]{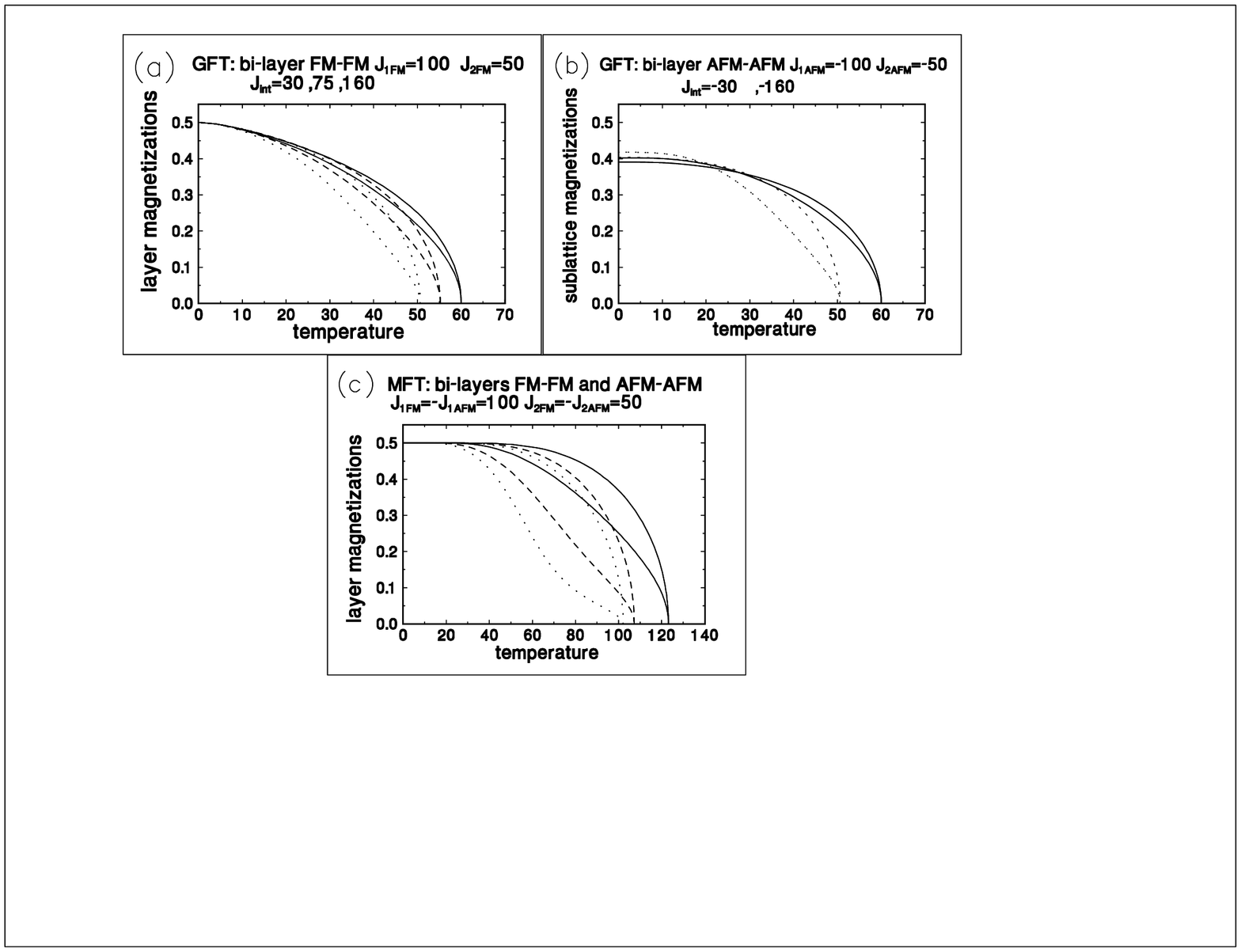}
\protect
\caption{ (a) Green's function theory (GFT) for the ferromagnetic bilayer: The
sublattice magnetizations are displayed as a function of the temperature
for different
interlayer couplings $J_{\rm int}= 30\ ({\rm dotted}),\ 75\ ({\rm
dashed}),\
160 \ ({\rm solid})$. The exchange interaction and anisotropy constants are
$J_{\rm 1FM}=100,\  J_{\rm 2FM}=50,\  D^z_{\rm 1FM}=1.0,\  D^z_{\rm
2FM}=1.0$.\\ (b) GFT for the antiferromagnetic bilayer: The sublattice
magnetizations are displayed as a function of the temperature for
two interlayer couplings $J_{\rm int}= -30\ ({\rm dotted}),\ -160\ ({\rm
solid} )$. The exchange interaction
and anisotropy constants
are $J_{\rm 1AFM}=-100,\ J_{\rm 2AFM}=-50,\ D^x_{\rm 1AFM}=-1.0,
D^x_{\rm 2AFM}=-1.0$.\\
(c) Mean field theory (MFT) for the ferromagnetic and antiferromagnetic
bilayers
with identical parameters:
$J_{\rm 1(2)FM}=|J_{\rm 1(2)AFM}|,\ $
$D_{\rm 1(2)FM}=|D_{\rm 1(2)AFM}|,\ $
$J_{\rm intFM}=|J_{\rm intAFM}|$.
} \label{fig1}
\end{center}
\end{figure}
\subsection{The coupled ferro- antiferromagnetic bilayer}
This is the most interesting case.
\begin{figure}[htb] \begin{center}
\protect
\includegraphics*[bb = 100  280 480 550,
angle=-90,clip=true,width=12cm]{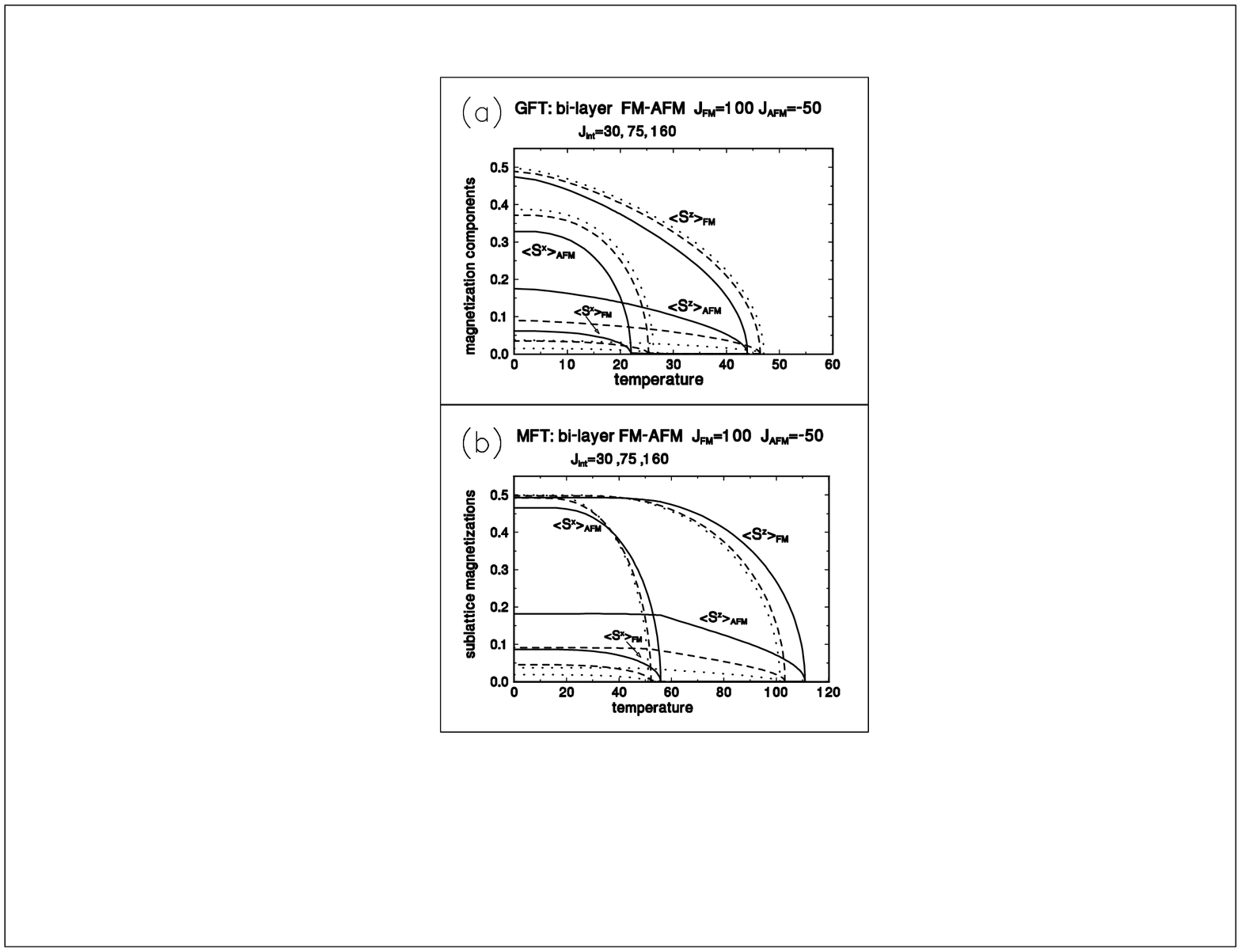}
\protect
\caption{ (a) Green's function theory(GFT): The
sublattice magnetizations of the ferro- and antiferromagnetic
sublattices are displayed as a function of the temperature for different
interlayer couplings $J_{\rm{int}}= 30,\ 75,\ 160$. The exchange interaction
and anisotropy constants are
$J_{\rm FM}=100,\ J_{\rm AFM}=-50,\ D^x_{\rm AFM}=-1.0,\ D^z_{\rm FM}=1.0$.\\
(b) Mean field theory (MFT): with the same parameters. }
\label{fig2}
\end{center}
\end{figure}
In the present study, we consider  two
in-plane magnetization components of each sublattice thus allowing noncollinear
magnetizations in both the FM and AFM layers. Our computer code, when
specialized to a single magnetization direction, reproduces
the results of reference \cite{MUH93}.
Without interlayer coupling,
the code also reproduces
 the results for both the monolayer ferromagnet and  monolayer antiferromagnet
simultaneously. The present choice of
anisotropies supports the orthogonal arrangement of the magnetizations of
the FM and AFM layers favoured by the exchange interaction alone.
The interlayer coupling destroys the perpendicular orientation of the
ferromagnet (in $z$-direction) with respect to the antiferromagnet (in
$x$-direction) even
at temperature $T=0$, as can be seen from figures 2 and 3. In figure 2(a) we show
the sublattice magnetizations of a sc lattice calculated with GFT for three
interlayer coupling strengths.
With a positive interlayer coupling all sublattice magnetizations develop a
positive $z$-component, whereas the $x$-components of the
two
sublattice magnetizations in each layer are oriented oppositely.
With increasing
temperature, all $x$-components decrease until they vanish at a common
temperature  $T^*_{\rm N\acute{e}el}$, slightly above the N\'{e}el
temperature
of the uncoupled AFM.  For $T>T^*_{\rm N\acute{e}el}$ all sublattice
magnetizations point
in the positive $z$-direction, the AFM layer assumes a
ferromagnetic arrangement, and remains so until a common critical temperature
$T_C$ is reached, at which the magnetic order vanishes altogether.
This is more clearly shown in figure 3 for the case of the strongest interlayer
coupling of figure 2. Due to the strong interlayer coupling, the magnetizations
of the layers are no longer collinear, even at $T=0$,
and turn more and more
into the $z$-direction with increasing temperature until
$T^*_{\rm N\acute{e}el}$, while the magnitudes of the magnetization vectors
shrink. Above $T^*_{\rm N\acute{e}el}$ the sublattice magnetizations stay
collinear until they vanish at the critical temperature $T_C$.
\begin{figure}[htb] \begin{center}
\protect
\includegraphics*[bb = 290  130 535 685,
angle=-90,clip=true,width=12cm]{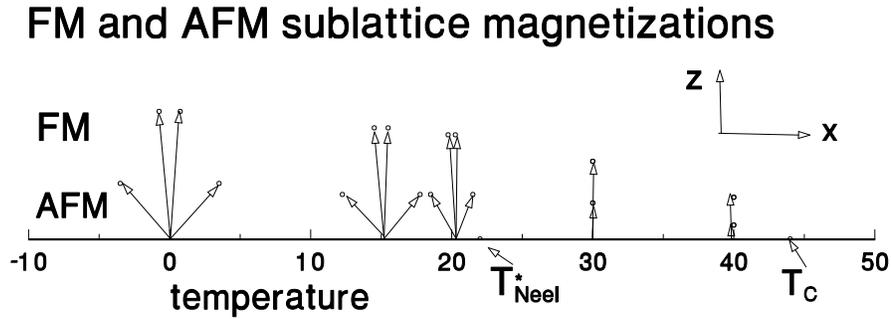}
\protect
\caption{The reorientation of the sublattice magnetizations of the
FM and AFM bilayer  as a function of the temperature
for the case
of the strongest coupling ($J_{\rm int}=160$) of figure 2.}
\label{fig3} \end{center} \end{figure}

With a negative interlayer coupling, the
antiferromagnetic sublattice spins rotate with increasing temperature into the
negative $z$-direction.
For $T>T^*_{\rm N\acute{e}el}$, the magnetizations of both layers point in
opposite directions, i.e. one has a
ferrimagnetic situation.

A remarkable behaviour can be seen in figure 2.  With increasing
interlayer coupling strength, the critical temperatures
$T^*_{\rm N\acute{e}el}= 26.24,\ 25.29,\ 22.00 $ and
$T_{\rm C}= 47.05,\ 46.34,\ 43.94 $ decrease in GFT.
This is in contrast to the behaviour in MFT
($T^*_{\rm N\acute{e}el}=51.19,\ 52.18,\ 55.86,\ $
$T_{\rm C}= 101.37,\ 103.28,\ 110.94$), where
the opposite is true (compare figures 2(a) and 2(b)). We attribute this
surprising
result to the  quantum fluctuations taken into account in
Green's function theory but neglected in a mean field treatment.
This behaviour depends, however, on the lattice type. It occurs for a sc
lattice, whereas
for fcc stacking an increase of $T_{\rm C}$ with increasing interlayer
coupling is obtained as in MFT. We show this by
deriving a formula for the critical temperature from GFT valid for both sc and
fcc bilayers. The magnetizations are
 collinearly arranged as $T\rightarrow T_{\rm C}$, i.e. the
x-components
of the magnetizations vanish and the set of Green's
functions reduces to two decoupled ($2\times 2$) problems from which one can
obtain via the spectral theorem two equations for the
magnetizations of both layers. Denoting the ratio of the magnetizations
by $\alpha_C=\la S^z_2\ra/\la S^z_1\ra$
these equations become in this limit:
\begin{equation}
2T_{{\rm C}}\frac{1}{N}\sum_{\bf k}\frac{a_2}
{a_1a_2-\alpha_C(\gamma_{\rm int}({\bf k})J_{\rm int})^2}=\frac{1}{2},
\label{26}
\end{equation}
\begin{equation}
2\alpha_CT_{\rm C}\frac{1}{N}\sum_{\bf k}\frac{a_2}
{a_1a_2-\alpha_C(\gamma_{\rm int}({\bf k})J_{\rm int})^2}=\frac{1}{2},
\label{27}
\end{equation}
with
\begin{eqnarray}
a_1&=&q_0(J_1+D_1^z)+\alpha_Cq_{\rm int}J_{\rm int}-\gamma_0({\bf
k})J_1,\nonumber
\\ a_2&=&\alpha_Cq_0(J_2+D_2^z)+q_{\rm int}J_{\rm int}-\alpha_C\gamma_0({\bf
k})J_2\ . \label{28}
\end{eqnarray}
For the notation see equations (\ref{8}-\ref{11a}).
The ${\bf k}$ summation is performed as an integral over the first Brillouin
zone, and $T_{\rm C}$ is determined from a selfconsistent solution
of the integral equations (\ref{26},\ref{27}).
Note that $T_{\rm C}$ is symmetric with respect to the sign of
$J_{\rm int}$, if simultaneously $\alpha_C\rightarrow -\alpha_C$.

In figure 4  the results for the critical temperatures
for sc and fcc stacking as a function of the interlayer coupling are shown.
In the upper panel we see that for a sc lattice $T_{\rm C}$ decreases
with increasing interaction strength in GFT, whereas it
increases with increasing interlayer interaction strength in MFT.
For fcc stacking, $T_{\rm C}$  increases with increasing interlayer interaction
strength for both GFT and MFT.
This behaviour we attribute to a different action of the quantum fluctuations
in sc and fcc stacking. For large interlayer couplings the critical
temperatures approach a saturation value in GFT, whereas they increase
infinitely in MFT, which cannot be the correct behaviour.

\begin{figure}[htb]
\begin{center}
\protect
\includegraphics*[bb = 90  180 480 560,
angle=-90,clip=true,width=14cm]{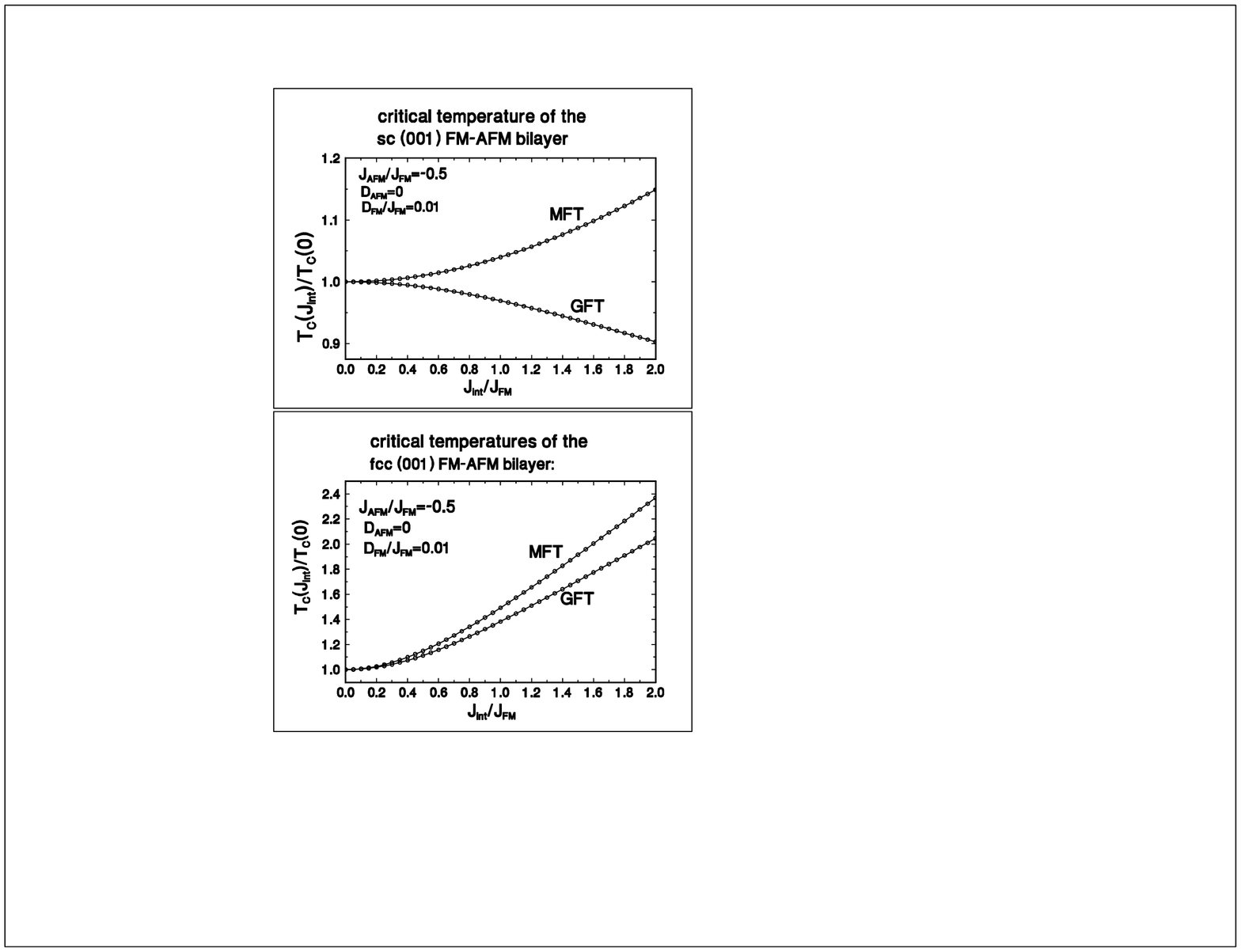}
\protect
\caption{ Critical temperatures $T_C(J_{\rm int})$ as
a function
of the interlayer coupling $J_{\rm int}$ of a simple cubic  (upper panel) and
a fcc (lower panel) FM-AFM bilayer ($T_{\rm C}$ is normalised
to $T_{\rm C}(J_{\rm int}=0)$). Results with mean field theory (MFT) and with
Green's function theory (GFT) are compared.
Note the qualitatively different behaviour of the
GFT results for sc and fcc stacking (see the text).}
\label{fig4}
\end{center}
\end{figure}

We now investigate the results of a sign change of the interlayer coupling
$J_{\rm int}$ for the case of an sc lattice. In figure 5 we display the
sublattice magnetizations for the ferromagnetic-
antiferromagnetic bilayer at temperature $T=0$ as functions of the interlayer
coupling $J_{\rm int}$. An asymmetric behaviour
of the magnetizations is observed with respect to the sign of $J_{\rm int}$.
The $z$-components of the sublattice magnetizations
are positive for positive interlayer couplings
and negative for negative couplings, while the $x$-components of the
magnetizations of the ferromagnetic sublattices interchange their roles.
More interesting is that the magnitudes of the corresponding components of the
sublattice magnetizations are not identical for $\pm J_{\rm int}$.
This situation with respect to the sign of the interlayer coupling does not
change at finite temperatures.
We discuss this without showing a figure. When the $x$-components of
the magnetizations vanish above the  temperature
$T^*_{\rm N\acute{e}el}$, the magnetizations of the
sublattices are collinear; for a positive interlayer coupling all
$z$-components
point in the positive $z$-direction, whereas for negative interlayer coupling
the magnetizations of the antiferromagnetic sublattices point in the negative
$z$-direction, i.e. opposite to the magnetizations of the ferromagnetic
sublattices.
The magnitudes of the magnetizations are larger for positive interlayer
coupling. Even though the magnetization components are somewhat different for
positive and negative interlayer couplings, numerical calculation and
inspection of equations (\ref{26},\ref{27}) show that the corresponding
critical temperatures are identical.
We attribute the observed asymmetries in the magnetizations in GFT to quantum
fluctuations in the non-collinear state, because there is no sensitivity
on the sign of the interlayer coupling in MFT \cite{JKD}.

\begin{figure}[htb] \begin{center}
\protect
\includegraphics*[bb = 50  60 570 730,
angle=-90,clip=true,width=9cm]{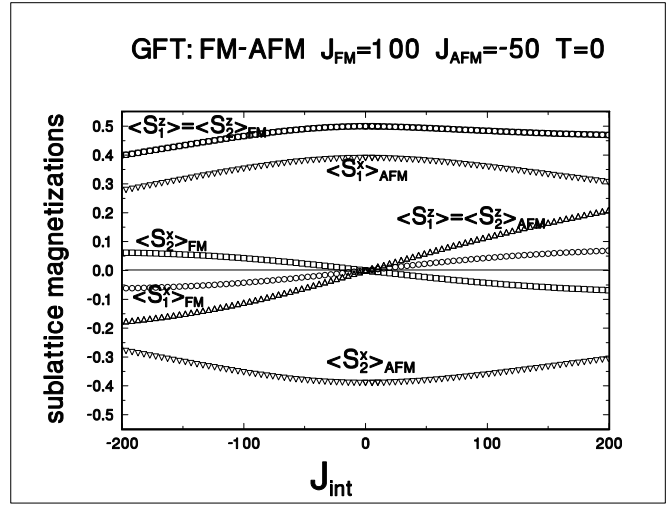}
\protect
\caption{Sublattice magnetizations for the ferro- antiferromagnetic bilayer as
function of the interlayer coupling strength $J_{\rm int}$. }
\label{fig5}
\end{center}
\end{figure}
\begin{figure}[htb] \begin{center}
\protect
\includegraphics*[bb = 50  60 570 730,
angle=-90,clip=true,width=9cm]{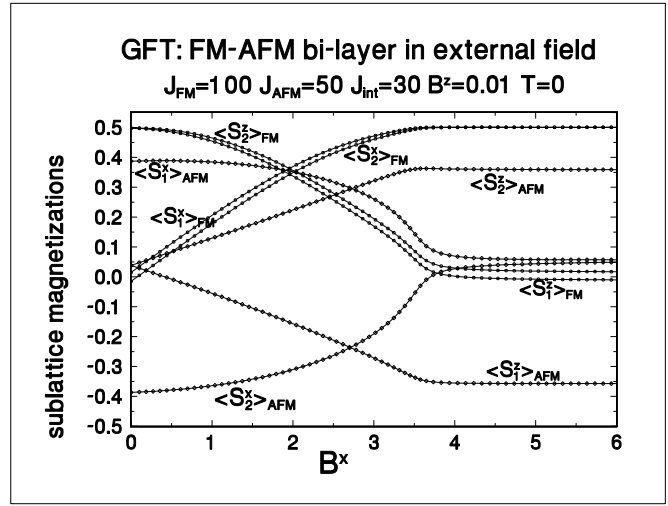}
\protect
\caption{Sublattice magnetizations for the ferro- antiferromagnetic bilayer as
funtion of a field in x-direction $B^x$ and a small field $B^z$ in z-direction.
} \label{fig6}
\end{center}
\end{figure}

Finally, we discuss the influence of an external
field on the
sublattice magnetizations. As the only example we show in figure 6 the
sublattice magnetizations at temperature $T=0$ as a function
of a field $B^x$ in the direction of the AFM easy axis
and a small perpendicular field component $B^z=0.01$.
At $B^x=0$, one has a nearly orthogonal configuration of the FM
($z$-direction) and AFM ($x$-direction) layers. Owing to the
positive interlayer coupling, both of the antiferromagnetic sublattice
magnetizations
have small positive $z$-components, and the ferromagnetic sublattice
magnetizations have a small positive and a negative $x$-component.
 With
increasing $B^x$, a field-induced magnetic reorientation occurs,
i. e. both of the $x$-components of the magnetization of the ferromagnet become
positive and increase, whereas the $z$-components decrease.
For very large fields, the magnetization of the ferromagnet
 points almost in the $x$-direction.
Accordingly, for a small $B^x$-field the AFM layer preserves its
nearly antiferromagnetic configuration  and the almost orthogonal magnetic
arrangement with respect to the ferromagnetic layer.
With increasing $B^x$, the AFM sublattice magnetizations
rotate more and more into the
$z$-direction, with
two small positive $x$-components.
The situation changes only slightly if one puts the $z$-component of the field
to zero, $B^z=0$. In this case the main difference is that the reorientation
transition  is sharp at a reorientation field $B^x=3.79$.
The behaviour of the field dependence at finite temperatures is analogous,
the only difference being that the magnetization components are reduced.

\section{Concluding remarks}

In the present paper we have developed a theory for coupling ferromagnetic and
antiferromagnetic Heisenberg layers in the framework of many-body Green's
function theory. The new feature is that we allow a non-collinear
orientation of the sublattice magnetizations for both of the ferromagnetic and
antiferromagnetic layers. We applied the theory to bilayer
systems with an in-plane orientation of the magnetizations. The theory yields
results which are in many instances qualitatively similar to results
previously obtained
by mean field theory \cite{JKD}. Owing to missing quantum fluctuations, MFT
cannot,
however, give the suppression of the magnetizations of the
antiferromagnetic
sublattices at low temperature. The main difference between MFT and GFT is, as
for the uncoupled
systems, the very different temperature scale.

An interesting effect is observed for the critical temperature $T_{\rm C}$ for
systems
with $J_{\rm FM}>|J_{\rm AFM}|$. For a sc lattice, $T_{\rm C}$ {\em
decreases} with
increasing interlayer coupling $J_{\rm int}$ in GFT, contrary to the behaviour
in
MFT, where the critical temperature {\em increases} with increasing $J_{\rm
int}$. For fcc
stacking, however, for both GFT and MFT the critical temperature
{\em increases}
with increasing $J_{\rm int}$. We attribute the different behaviour of the GFT
results to a different action of the quantum fluctuations for sc and fcc
stacking.

In the future we will study multilayer systems with mixed FM and AFM couplings.
The code is written in such a
way that it also allows the description of the coupling of ferromagnetic layers
by an antiferromagnetic interlayer coupling and vice versa. Because we have
already
included magnetic fields, the theory may also be the basis for studying the
exchange bias effect, where it seems, however, to be necessary to include
interface disorder \cite{EBdis} in some way, for instance by introducing more
sublattices per layer with different magnetic arrangements.

\end{document}